\documentclass{elsart}

\usepackage{graphicx}
\usepackage{amssymb}

\begin{document}

\begin{frontmatter}

\title{Different regimes of synchronization in nonidentical
time-delayed maps}

\author[M]{Cristina Masoller},
\ead{cris@fisica.edu.uy}
\author[Z]{Dami\'an H. Zanette}
\ead{zanette@cab.cnea.gov.ar}

\address[M]{Instituto de F\'{\i}sica, Facultad de Ciencias,
Universidad de la Rep\'ublica, Igu\'a  4225, Montevideo 11400,
Uruguay}
\address[Z]{Consejo Nacional de Investigaciones Cient\'{\i}ficas y
T\'ecnicas, Centro At\'omico Bariloche and Instituto Balseiro,
8400 Bariloche, R\'{\i}o Negro, Argentina}

\begin{abstract}
We study the synchronization of time-delayed nonidentical maps
subject to unidirectional (master-slave) coupling. The individual
dynamics of the maps have a delay $n_1$, and the coupling acts
with a delay $n_2$. We show analytically that, suitably tuning
the slave map parameters, two distinct synchronization regimes can
occur. In one regime the lag time between the slave and the
master maps is given by the delay of the coupling, $n_2$, while
in the other regime is given by the difference between the
delays, $n_1-n_2$. We analyze the effect of the coupling strength
on the different synchronization regimes in logistic and
H\'{e}non maps.

\end{abstract}

\end{frontmatter}

\section{Introduction}

Delay differential equations have received much attention over
the years because of the significant role of delayed feedback in
the dynamics of many physical and biological systems  \cite{a}.
On the one hand, delay-differential systems often exhibit
multistability --i.e., the coexistence of several attractors-- and
multistability enables such systems to act as memory devices
\cite{b,c}, an idea first suggested by Ikeda and Matsumoto
\cite{d}. On the other, the study of delay systems is motivated
by the fact that these systems exhibit high-dimensional chaos and
therefore can be used in communication systems based on chaotic
synchronization, to securely encrypt information into their
chaotic outputs \cite{e,f}.

Several authors have recently shown the existence of two
different regimes of synchronization in nonidentical coupled
time-delayed differential equations, in a master-slave
configuration \cite{1,2,3,4,5}. When the master and the slave
systems have both the same amount of delayed intrinsic feedback
and the delay time $\tau$ is the same for both systems, the slave
system variables synchronize with the master system variables
with a lag time that is equal to the delay of the coupling,
$\tau_c$. In other words, $x_2(t)= x_1(t-\tau_{c})$, where
$x_1(t)$ and $x_2(t)$ represent the states of the master and
slave systems respectively.

On the other hand, when the master and the slave systems have the
same amount of delayed intrinsic feedback and the feedback
coefficient of the master system is equal to the sum of the
feedback coefficient of the slave system and the coupling
coefficient, the systems synchronize with a different lag-time.
In this case $x_2(t)= x_1(t+\tau-\tau_{c})$. Two synchronization
regimes are therefore possible for appropriate choices of the
slave map parameters, as shown analytically in specific instances
\cite{5,6}.

In this paper we show that these two synchronization regimes can
also occur in time-discrete dynamical systems. In Section 2 we
consider generic nonidentical delayed maps, unidirectionally
coupled, and define the two synchronization regimes. As an
illustration, in Section 3 we analyze logistic and H\'{e}non
delayed maps. Finally, Section 4 presents our conclusions.

\section{Master-slave coupled delayed maps}

We consider a generic master map of the form
\begin{equation} \label{master}
x_{n+1}= \alpha f(x_n) + \beta f(x_{n-n_1}) + g(x_n),
\end{equation}
where $\alpha$ and $\beta$ are parameters. The slave map is given
by
\begin{equation} \label{slave}
y_{n+1}= \alpha_s f(y_n) + \beta_s f(y_{n-n_1})+ g(y_n)  + \eta
f(x_{n-n_2}).
\end{equation}

If the parameters of the slave map are tuned in such a way that
$\alpha_s=\alpha$ and $\beta_s=\beta-\eta$ (case I),  the slave
map reduces to
\begin{equation}
y_{n+1}=  \alpha f(y_n) + \beta f(y_{n-n_1}) + g(y_n)
 + \eta [f(x_{n-n_2})- f(y_{n-n_1})]. \label{lag1}
\end{equation}
Full synchronization can be expected for sufficiently large $\eta$
on the synchronization manifold $y_n=x_{n+n_1-n_2}$. On the other
hand, if the parameters of the slave map are tuned in such a way
that $\alpha_s = \alpha - \eta $  and $\beta_s=\beta$ (case II),
the slave map reads
\begin{equation}
y_{n+1}=  \alpha f(y_n) + \beta f(y_{n-n_1})+ g(y_n) + \eta
[f(x_{n-n_2})- f(y_n)], \label{lag2}
\end{equation}
and full synchronization may occur on the synchronization manifold
$y_n=x_{n-n_2}$. Therefore, depending on the parameters of the
slave map, $\alpha_s$ and $\beta_s$, full synchronization can
take place with two different lag times, $\Delta_{\rm I}= n_2-n_1$
in the case of Eq. (\ref{lag1}) and $\Delta_{\rm II} =n_2$ in the
case of Eq. (\ref{lag2}). Note that, in  case I, one can have
anticipated syncronization for $n_1>n_2$ [refs. de anticipated
synchronization].

The actual possibility of observing full synchronization in
either case is determined by the stability of the synchronized
state. Linear stability analysis of Eqs. (\ref{lag1}) and
(\ref{lag2}) can be carried out by   noticing first that the two
equations can be written in a unified form as
\begin{equation}
y_{n+1}=  h(y_n) + \beta f(y_{n-n_1})  + \eta [f(x_{n-n_2})-
f(y_{n-n_3})], \label{gen}
\end{equation}
with $h(y)=\alpha f(y)+g(y)$. In Eq. (\ref{lag1}) we have
$n_3=n_1$, whereas in Eq. (\ref{lag2}) we have $n_3=0$. The
synchronization manifold is $y_n=x_{n+n_3-n_2}$. Applying a
perturbation $y_n=x_{n+n_3-n_2}+\delta_n$, replacing in Eq.
(\ref{gen}), and taking into account Eq. (\ref{master}) we get, to
the first order in the perturbation,
\begin{equation}
\delta_{n+1}=h'(x_{n+n_3-n_2})\delta_n+\beta
f'(x_{n+n_3-n_2-n_1}) \delta_{n-n_1}-\eta f'(x_{n-n_2})
\delta_{n-n_3}, \label{delta}
\end{equation}
where primes indicate derivatives. Equation (\ref{delta}) can be
formally integrated by introducing a linear $(N+1)$-dimensional
map, with $N= \max\{ n_1,n_3 \}$, for a variable ${\bf r}_n=
(r^0_n, r^1_n,\dots,r^{N}_n)$, with $r^k_n= \delta_{n-k}$. This
equivalent map is given by
\begin{equation}
{\bf r}_{n+1}= M_n {\bf r}_n, \label{z}
\end{equation}
where the elements of the matrix $M_n$ are given by the
(time-dependent) coefficients in Eq. (\ref{delta}) \cite{nos}. The
solution to Eq. (\ref{z}) reads
\begin{equation} \label{U}
{\bf r}_n = U_n {\bf r}_0= M_{n-1}M_{n-2} \cdots M_1M_0 {\bf r}_0,
\end{equation}
so that the state of full synchronization is linearly stable if
all the eigenvalues of the evolution matrix $U_n$ vanish for $n
\to \infty$. Whether this condition holds or not for a given
value of the coupling constant $\eta$ can be readily verified by
numerical means. Note that all the elements of matrix $M_n$,
given by the coefficients in Eq. (\ref{delta}), involve a delay
$n_2$ which thus acts as a uniform time shift. This fact implies
that, in the limit $n\to \infty$, the eigenvalues of $U_n$ become
independent of $n_2$. Consequently, the value of $n_2$ is
irrelevant to the stability of full synchronization (cf.
\cite{nos}). Note carefully that, generally, full synchronization
for the two cases considered above will be stable on two different
ranges of the coupling constant $\eta$. If the master map is
chaotic, we expect that for sufficiently small and large $\eta$
full synchronization is unstable and stable in both cases,
respectively, while for intermediate values only one of the cases
admits full synchronization.

\begin{figure}
\centering \resizebox{.6\columnwidth}{!}{\includegraphics{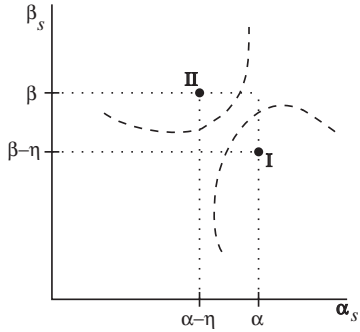}}
\caption{Schematic representation in the parameter space
($\alpha_s$, $\beta_s$) of the regions where synchronization with
lag time $\Delta_{\rm I}$ and $\Delta_{\rm II}$ occur.}
\label{f0}
\end{figure}

We have seen above that full synchronization is possible at two
points in the ($\alpha_s$,$\beta_s$) parameter space, with
different lag times in each case. To encounter such synchronized
states, the slave-map parameters must be exactly tuned on one of
those synchronization points. Their location is schematically
shown in Fig. \ref{f0}. Though when the slave system is slightly
detuned with respect to the synchronization points full
synchronization will not occur, it is expected that the slave-map
orbit follows approximately the master-map orbit with the same lag
time, $\Delta_{\rm I}$ or $\Delta_{\rm II}$. To quantitatively
characterize the degree of synchronization between the two orbits
and the respective lag time, we may use the so-called similarity
function $S_\Delta$, defined as \cite{5}
\begin{equation}
S_\Delta^2={{\langle [ x_{n+\Delta}-y_n]^2\rangle } \over{[
\langle x_n^2\rangle \langle y_n^2\rangle ]^{1/2}}} ,
\end {equation}
where the brakets $\langle \cdot \rangle$ stand for time averages
over asymptotically large times. If $x_n$ and $y_n$ are
independent time series with similar mean value and dispersion we
have $S_\Delta \approx \sqrt{2} \approx 1.4$. If, on the other
hand, there is full synchronization with lag time $\Delta$,
$S_\Delta=0$. The similarity function $S_\Delta$ can be
determined, at each point ($\alpha_s$,$\beta_s$) parameter space
and for each lag time $\Delta$. At $(\alpha,\beta-\eta)$ we
should have $S_{\Delta_{\rm I}}=0$, while at $(\alpha-\eta,
\beta)$ we should have $S_{\Delta_{\rm II}}=0$. It is expected,
moreover, that in a region around each synchronization point the
similarity function attains a minimum, as a function of the lag
time, for $\Delta= \Delta_{\rm I}$ and $\Delta=\Delta_{\rm II}$,
respectively. These regions are qualitatively illustrated in Fig.
\ref{f0}. In the remaining of the parameter space,  as far as the
slave-map orbits do not diverge, the similarity function can
attain a minimum for any other value of $\Delta$ --without
reaching, however, $S_\Delta=0$. Note that the boundaries of such
regions will in general depend on the coupling constant $\eta$.
In the following sections, we study these aspects of
synchronization in Eqs. (\ref{master}) and (\ref{slave}) for
logistic and H\'{e}non delay maps in their chaotic regime.

\section{Synchronization of delayed logistic and H\'{e}non maps}

\subsection{Logistic maps}

As a first illustration of the synchronization properties of Eqs.
(\ref{master}) and (\ref{slave}) in cases I and II, we consider
the choice $f(x)=x(1-x)$ and $g(x)=0$. The master system becomes a
delayed logistic map
\begin{equation} \label{mlog}
x_{n+1}= \alpha x_n(1-x_n) + \beta x_{n-n_1} (1-x_{n-n_1}),
\end{equation}
whose orbits are bounded to the interval $(0,1)$ for $0<\alpha
,\beta$ and $\alpha+\beta \le 4$. In different regions of the
parameter space $(\alpha,\beta)$ and depending on the delay
$n_1$, this system displays periodic, quasiperiodic and chaotic
evolution. The corresponding slave map is given by
\begin{equation} \label{slog}
y_{n+1}= \alpha_s y_n(1-y_n) + \beta_s y_{n-n_1}(1-y_{n-n_1}) +
\eta x_{n-n_2} (1-x_{n-n_2}).
\end{equation}
Its orbits are nondivergent for $0<\alpha_s ,\beta_s,\eta$ and
$\alpha_s+\beta_s+\eta<4$.

Figure 2 displays the synchronization regions in the parameter
space ($\alpha_s$, $\beta_s$) and how they vary as the coupling
coefficient $\eta$ increases. The master map parameters are
$\alpha=1.8$, $\beta=2.1$ and the delay times are $n_1=2$,
$n_2=3$. For each pair ($\alpha_s$, $\beta_s$) we have determined
$S_\Delta$ as a function of the lag-time $\Delta$, and detected
the value of $\Delta$ for which the similarity function attains
its minimum $\min (S_\Delta)$. The left column of Fig. 2 displays
this minimum  for three values of $\eta$. Light tones represent
low values of $\min(S_\Delta)$, i.e. high master-slave
correlation, while darker tones correspond to poor correlation
[$\min(S_\Delta) \sim 1$]. In the black upper-right region the
slave-map orbits diverge.

The right column of Fig. 2 displays the lag-time $\Delta$ at
which the similarity function attains its minimum. The region
where $S_\Delta$ is minimal with lag-time $\Delta_{\rm
I}=n_2-n_1=3$ is represented by the darker gray tone, while the
region with lag-time $\Delta_{\rm II}=n_2=3$ is represented by the
lighter gray tone. White represents the parameter region where
the minimum value of the similarity function occurs for a
lag-time which is different from $\Delta_{\rm I}$ or $\Delta_{\rm
II}$. Black represents the parameter region where the trajectory
of the slave map diverges.

For low coupling intensity, $\eta=0.2$, both synchronization
regimes are unstable [Figs. 2 (a, b)]. The synchronization regions
are not well defined and have fuzzy boundaries. While the minimum
of the similarity function at point I occurs at the expected
lag-time $\Delta_{\rm I}$, the minimum of the similarity function
at point II occurs at a different lag-time. Notice, in fact, that
in Fig. 2 (b) point II is in the white region that represents a
lag-time different from $\Delta_{\rm I}$ or $\Delta_{\rm II}$. The
region corresponding to each regime is disconnected and quite
complex in shape, with parts in distant zones of the parameter
space. Note, for instance, the light-gray zones near $\beta_s=0$
where master-slave correlation is however rather poor.

As the coupling intensity grows, zones I and II become more
uniform and increase in total extension. For $\eta=0.8$ only
regime I is stable [Figs. 2 (c, d)]. In this case,
$\min(S_\Delta)$ at points I and II is equal to 0 and 0.2
respectively. For large enough $\eta$, both regimes are stable
[Figs. 2 (e, f); $\eta=1.2$] and, as expected, $\min(S_\Delta)$
equals zero at points I and II.

\begin{figure}
\centering \resizebox{.8\columnwidth}{!}{\includegraphics{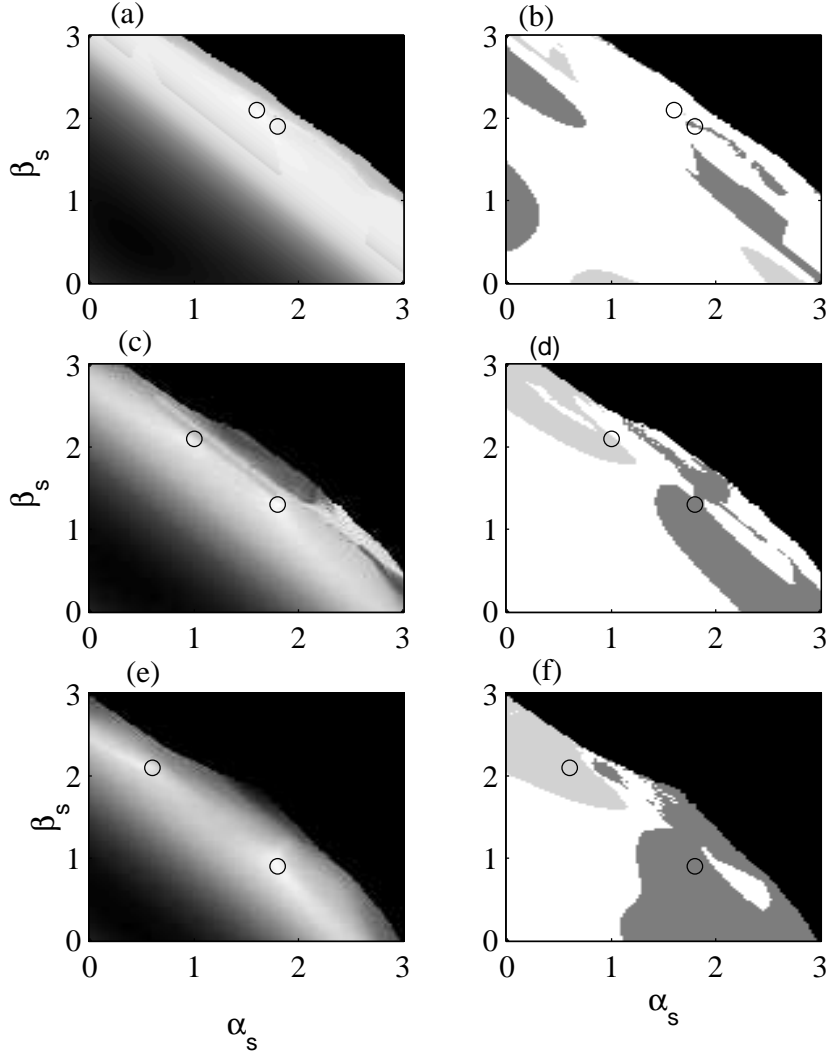}}
\caption{Synchronization regions in the ($\alpha_s$,$\beta_s$)
parameter space in the case of the logistic map, for increasing
coupling: (a, b) $\eta=0.2$; (c, d) $\eta=0.8$; (e, f)
$\eta=1.2$. The master map parameters are $\alpha=1.8$,
$\beta=2.1$, and the delay times are $n_1=2$ and $n_2=3$. The
right column displays the minimum of the similarity function.
Light tones represent low values of $\min(S_\Delta)$ (good
master-slave correlation) and {\it vice versa}. Black represents
the region where the slave-map trajectories diverge. The left
column displays the lag-time where the minimum value of
$S_\Delta$ occurs. In the light-gray region, the lag-time is
$\Delta_{\rm II}=n_2=3$, while in the dark-gray region it is
$\Delta_{\rm I}=n_2-n_1=1$. In the white region the lag-time is
different from $\Delta_{\rm I}$ or $\Delta_{\rm II}$. The small
circles stand at the synchronization points I and II,
($\alpha_s=\alpha$, $\beta_s=\beta-\eta$) and
($\alpha_s=\alpha-\eta)$, $\beta_s=\beta$), respectively. }
\end{figure}

Figure 3 illustrates the master-slave correlation at different
points of parameter space for $\eta=1.0$ (all other parameters
are as in Fig. 2). In this case, type I synchronization is stable
[Fig. 3 (a)] but type II is not [Fig. 3 (b)], and it is worth
mentioning that the minimum value of the similarity function,
$S_\Delta=0.057$, does not occur at point II ($\alpha_s=0.8$,
$\beta_s=2.1$) but at a point close to it ($\alpha_s=0.75$,
$\beta_s=2.1$). Figure 3 (c) displays the correlation plot at a
point where the lag-time at which $S_\Delta$ attains its minimum
is $\Delta=-15$.

\begin{figure}
\centering \resizebox{.8\columnwidth}{!}{\includegraphics{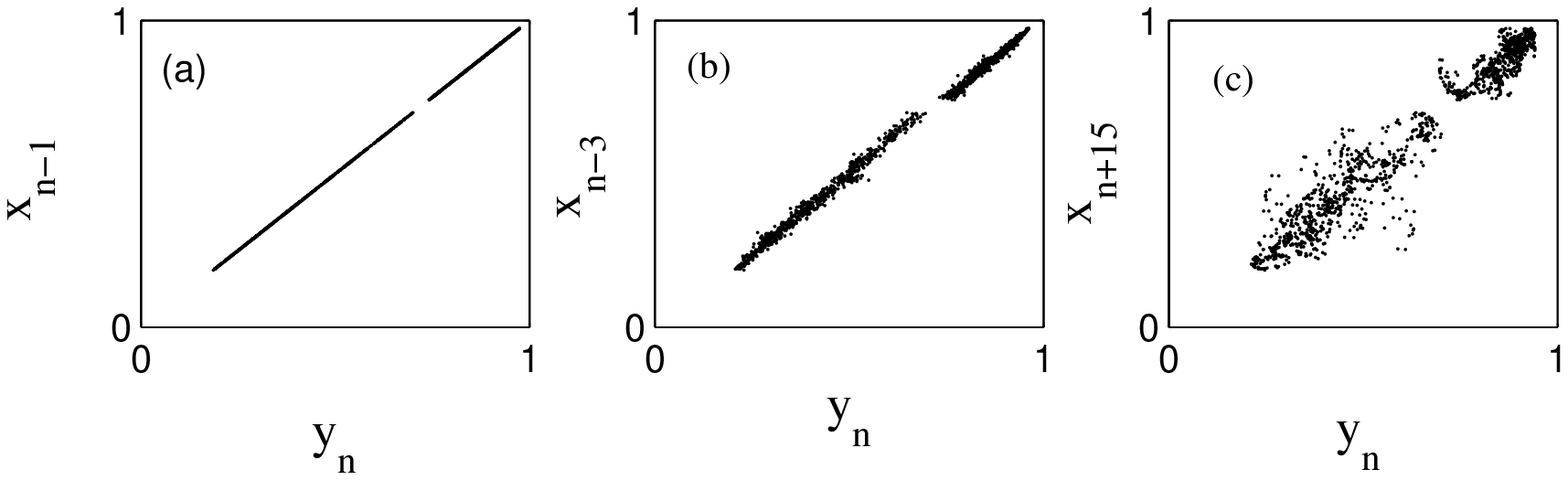}}
\caption{Correlation plots for $\eta=1.0$, $n_1=2$, $n_2=3$ and (a)
$\alpha_s=1.8$, $\beta_s=1.1$, $\min(S_\Delta)=0$; (b)
$\alpha_s=0.75$, $\beta_s=2.1$, $\min(S_\Delta)=0.057$; (c)
$\alpha_s=1.25$, $\beta_s=1.525$, $\min(S_\Delta)=0.25$. The
master-map parameters are as in Fig. 2. }
\end{figure}

\subsection{H\'{e}non maps}

As a second example, we study now a master-slave configuration
where each element is a two-dimensional delay map, namely, a
H\'{e}non-like map. The evolution of the master coordinates ${\bf
x }_n = (u_n,v_n) $ is given by the functions ${\bf f} ({\bf x})=
(-u^2,0)$ and ${\bf g} ({\bf x})= (1+v,bu)$, so that the master
system is
\begin{equation}
\begin{array}{ll}
u_{n+1} &= 1 - \alpha u^2_n - \beta u^2_{n-n_1} + v_n,
\\ v_{n+1} &=  b u_n,
\end{array}
\end{equation}
cf.  Eq. (\ref{master}). In the following we choose $b=0.3$. The
slave system, with coordinates ${\bf y }_n = (w_n,z_n)$, is
governed by the equations
\begin{equation} \label{x}
\begin{array}{ll}
w_{n+1} &= 1 - \alpha_s w^2_n - \beta_s w^2_{n-n_1}+ z_n -\eta
u^2_{n-n_2},  \\ z_{n+1} &= b w_n,
\end{array}
\end{equation}
so that coupling acts on the first coordinate only. The
synchronization manifold is given by
\begin{equation}
\begin{array}{ll}
w_n &= u_{n+n_3-n_2},
\\
z_n  &= v_{n+n_3-n_2},
\end{array}
\end{equation}
where, as before, $n_3=n_1$ in case I and $n_3=0$ in case II.

Next we study in which regions of parameters the different
synchronization regimes occur. We take parameters for the master
map $\alpha$, $\beta$, $n_1$ such that its dynamics is chaotic.

Figure 4 displays the minimum of the similarity function and the
lag-time for which the minimum occurs, in the parameter space
($\alpha_s, \beta_s$). The results are similar to those found
with the logistic map. For weak coupling the synchronization
regions are not well defined, but as the coupling increases their
size grows and the boundary between them becomes well defined. For
large $\eta$ both synchronization regimes  are stable. Thus, it
can be though that a small variation of the slave map parameters
$\alpha_s$ or $\beta_s$ near the boundary region might induce a
transition from synchronization with lag-time $\Delta_{\rm I}$ to
synchronization with lag-time $\Delta_{\rm II}$ or viceversa.
However, near the boundary region we find $\min(S_\Delta) \approx
0.5$, which indicates bad synchronization. Therefore, while the
lag-time at which the minimum value of $S_\Delta$ occurs changes
abruptly (from $\Delta_{\rm I}$ to $\Delta_{\rm II}$), there is no
sharp transition between one regime of synchronization to the
other. If   the slave map parameters are gradually modified from
point I to point II, synchronization with lag-time $\Delta_{\rm
I}$ is gradually lost, and as we enter the region II,
synchronization with lag-time $\Delta_{\rm II}$ is gradually
established.

\begin{figure}
\centering
\resizebox{.8\columnwidth}{!}{\includegraphics{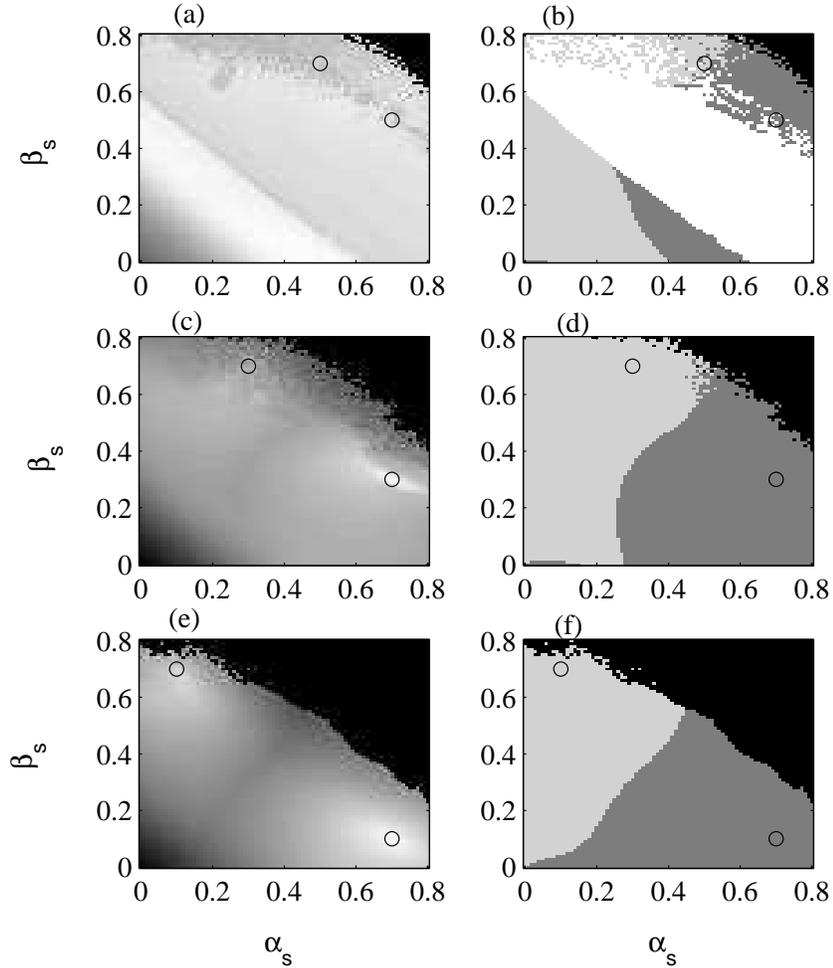}}
\caption{ Synchronization regions for delayed H\'{e}non maps. The parameters
of the master map are $\alpha=0.7$, $\beta=0.7$, and delay times
are $n_1=6$ and $n_2=3$. The left column displays the minimum of
the similarity function and the right column displays the
lag-time where $\min(S_\Delta)$ occurs. In the light-gray region,
the lag-time is $\Delta_{\rm II}=n_2=3$, while in the dark-gray
region it is $\Delta_{\rm I}=n_2-n_1=-3$. In the white region of
(b) the lag-time is different from $\Delta_{\rm I}$ or
$\Delta_{\rm II}$. The small circles indicate the points
($\alpha_s=\alpha-\eta$, $\beta_s=\beta$), and
($\alpha_s=\alpha$, $\beta_s=\beta-\eta$. (a, b) $\eta=0.2$;
(c,d) $\eta=0.4$; (e, f) $\eta=0.6$. }
\end{figure}

\section{Conclusion}

We have studied two regimes of synchronization of delayed
nonidentical maps. We have shown analytically that, by suitably
tuning the slave map parameters, two distinct synchronization
regimes can occur. In one regime the lag time between the slave
and the master maps is given by the delay of the coupling, $n_2$,
while in the other regime is given by the difference between the
delays, $n_1-n_2$. We have also shown that these two regimes are
actually two particular cases of synchronization with a slave map
that is identical to the master map but that has two delayed
feedback terms.

The two synchronization regimes has been exemplified by
considering delayed logistic and H\'{e}non maps. In both cases,
the synchronization regimes are simultaneously stable only for
large values of the coupling $\eta$, and therefore, they occur at
parameters of the slave map, ($\alpha_s$,$\beta_s$), which are
far away from each other. In other words, our results show that
in the case of delayed logistic and H\'{e}non maps, a small
variation of a parameter of the slave map can not induce a
transition from regime I to regime II or viceversa, since they
occur in distant regions of the parameter space. On the contrary,
in the case of semiconductor lasers with optical feedback, it has
been shown numerically \cite{2,4} that close to the lasing
threshold, by carefully tuning a parameter of the slave laser one
can induce a transition from one regime of synchronization to the
other. It will be interesting to study a delayed map which shows
this type of transition, and allows an analytical investigation
of the phenomenon.

\section*{Acknowledgements}

This work was supported by Proyecto de Desarrollo de Ciencias
B\'asicas (PE\-DE\-CI\-BA) and by Comisi\'on Sectorial de
Investigaci\'on Cient\'{\i}fica (CSIC), U\-ru\-guay.

\end{document}